\documentclass[letterpaper,aps,preprint,nofootinbib]{revtex4}%
\usepackage{amsfonts}
\usepackage[utf8]{inputenc}
\usepackage{amsmath}
\usepackage{amssymb}
\usepackage{float}
\usepackage{graphicx}%
\providecommand{\U}[1]{\protect\rule{.1in}{.1in}}

\begin{document}
\title{Critical points of the Exotic Massive 3D Gravity}
\author{Gaston Giribet$^1$, Andr\'es Goya$^1$, Edmundo Lavia$^{2,3}$, Julio Oliva$^4$}
\affiliation{$^{1}$Physics Department, University of Buenos Aires \& IFIBA-CONICET,\\Ciudad Universitaria, pabell\'on 1, 1428, Buenos Aires, Argentina}
\affiliation{$^{2}$Acoustic Propagation Department. Argentinian Navy Research Office (DIIV), \\Laprida 555, 1638 Vicente L\'opez, Buenos Aires, Argentina}
\affiliation{$^{3}$UNIDEF (National Council of Scientific and Technical Research – Ministry of Defense)}
\affiliation{$^{4}$Physics Department, University of Concepci\'on, Casilla 160-C, Concepci\'on, Chile\footnote{gaston@df.uba.ar, af.goya@df.uba.ar, edmundolavia@fibertel.com.ar, juoliva@udec.cl}}

\begin{abstract}
Exotic Massive 3D Gravity (EMG) is a higher order generalization of Topologically Massive 
Gravity. As in other theories of this sort, the conserved charges associated to the asymptotic diffeomorphisms that 
preserve the boundary conditions in AdS$_3$ spacetime span two copies of the Virasoro algebra with non-vanishing central 
charges. Here, we discuss the values of these central charges and the corresponding conformal anomaly in relation to the 
phase space of the theory.
\end{abstract}
\maketitle

\section{Introduction}
\label{sec:intro}

Chiral Gravity (CG) \cite{Chiral.Gravity} is a parity-odd theory of massive gravity in 3 dimensions around Anti-de 
Sitter (AdS) space whose mass parameter takes a value such that either the left-moving or the right-moving central 
charge of the dual conformal field theory (CFT) vanishes. The vanishing of one of the central charges is usually 
associated to the emergence in the bulk of a massless graviton mode, which produces a long range interaction 
characterized by a logarithmic fall-off near the boundary \cite{Grumiller.Johansson:TMG.log.conditions, GAY:0811}
\begin{equation}
h\sim \log (r) \,.\label{loga}
\end{equation}
This log-mode has negative energy in the bulk, and it makes the dual CFT$_2$ to be non-unitary. This is the reason why, 
in order to define CG in a consistent way, one needs to impose strong asymptotically AdS$_3$ boundary conditions that 
suffice to eliminate modes like (\ref{loga}) \cite{Chiral.Gravity.Log}. These boundary conditions are 
the Brown-Henneaux asymptotic conditions, i.e. the same as in general relativity (GR) \cite{Brown.Henneaux:Central.Charge}. If such boundary conditions are imposed, then the dual theory turns out to be a chiral CFT$_2$.

CG was originally formulated as a particular case of Topologically Massive Gravity (TMG) \cite{TMG, TMG.PRL} with 
negative cosmological constant. However, it can be easily generalized by adding to the TMG field equations other 
contributions, also representing sensible massive deformations of Einstein equations, such as New Massive Gravity (NMG) 
\cite{NMG, More.NMG}, Minimal Massive Gravity (MMG) \cite{MMG}, or the recently proposed Exotic Massive Gravity (EMG) 
\cite{EMG}, all these being particular cases of a more general set of models \cite{Third.Way.Gravity}. In a series of 
recent papers \cite{Chernicoff:EMG.Vacua, Giribet.Oliva:EMG.More.Vacua, MOS:EMG.Energy, 
BMT:3D.mass.grav.Asymptotic.charges}, EMG coupled to TMG around AdS$_3$ was studied and the properties of its dual CFT$_2$ were 
analyzed (see also \cite{Bachian, Ozkan:2019iga, Kilicarslan:2019ply, Afshar:2019npk}). In particular, the values of the central charges were obtained and the special features the theory exhibits when those charges vanish were studied. This is the problem we want to revisit here.

\section{Exotic Massive Gravity}

In the metric formalism\footnote{The theory also admits a Chern-Simons like formulation in terms of the vielbein $e^a_{\mu}$ and the spin connection $\omega^{ab}_{\mu}$; see \cite{EMG, BMT:3D.mass.grav.Asymptotic.charges} for details.}, EMG is defined by the following field equations \cite{EMG}
\begin{equation}
\label{eq:eom}
	R_{\mu \nu}-\dfrac{1}{2} R g_{\mu \nu} + \Lambda g_{\mu \nu} + \dfrac{1}{\mu} C_{\mu \nu} = T_{\mu \nu} \,,
\end{equation}
where the Cotton tensor is
\begin{equation}
\label{eq:Cotton}
	C_{\mu \nu} = \dfrac{1}{2} \varepsilon_{\mu}^{\:\:\: \alpha \beta} \nabla_{\alpha} \left(R_{\beta \nu}-\frac{1}{4}g_{\beta \nu} R \right) + \dfrac{1}{2} \varepsilon_{\nu}^{\:\:\: \alpha \beta} \nabla_{\alpha} \left( R_{\beta \mu}-\frac{1}{4}g_{\beta \mu}R \right) \,,
\end{equation}
and where
\begin{equation}
\label{eq:HL.tensors}
	T_{\mu \nu} = \dfrac{1}{m^{2}} 
	\varepsilon_{\mu}^{\:\:\: \alpha \beta} \nabla_{\alpha} C_{\beta \nu}
	- \dfrac{1}{2m^{4}} 
	\varepsilon_{\mu}^{\:\:\: \alpha \beta} \varepsilon_{\nu}^{\:\:\: \gamma \sigma} C_{\alpha \gamma} C_{\beta \sigma} \,.
\end{equation}

The limit $m\to\infty$ of this theory leads to TMG, and the limit $\mu\to 0$ gives the 3D conformal gravity. For $|\mu |<\infty $ the theory does not have a definite parity since while GR and the Exotic terms are parity-even, the Cotton tensor is parity-odd. 

Equations (\ref{eq:eom}) do not follow from a variational principle as they are covariantly conserved only on-shell; see 
\cite{Third.Way.Gravity} for details about this mechanism.

As in the case of other massive deformations of 3-dimensional Einstein theory, the conserved charges associated to the 
asymptotic diffeomorphisms in AdS$_3$ span two copies of the Virasoro algebra \cite{Brown.Henneaux:Central.Charge}; namely 
\begin{equation}
[L^{\pm}_m , L^{\pm}_n] = (m-n)L^{\pm }_{m+n}+\frac{c_{\pm}}{12}m(m^2-1)\delta_{m+n,0} \,,
\end{equation}
with $[L^{+}_m , L^{-}_n] =0$. The central charges $c_{\pm}$, according to the computation of \cite{BMT:3D.mass.grav.Asymptotic.charges}, 
are given by
\begin{equation}
\label{eq:central.charges}
	c_{\pm} = \dfrac{3 \ell}{2 G}\left[ -\dfrac{\ell m^{2}}{\mu} \pm\left(1+\dfrac{m^{2}}{\mu^{2}}-\dfrac{1}{\ell^2 m^2}\right) \right] \,,
\end{equation}
where $\ell = 1/\sqrt{-\Lambda }$ is the radius of AdS$_3$. This result for $c_{\pm }$ differs from the one obtained in \cite{Giribet.Oliva:EMG.More.Vacua}. 
In particular, the charges (\ref{eq:central.charges}) exhibit two set of critical points; namely
\begin{equation}
\label{eq:chiral.curves}
	\mu_{\mathrm{crit}, 1\pm } = \pm \, m^{2}\ell \,, \quad \quad \mu_{\mathrm{crit}, 2\pm } = \pm \dfrac{ m^{2} \ell}{m^{2}\ell^{2} -1} \,,
\end{equation}
where either $c_{-}$ or $c_{+}$ vanishes; see figure 1. In \cite{Giribet.Oliva:EMG.More.Vacua}, in contrast, only the points 
$\mu_{\mathrm{crit}, 2\pm }$ were identified as critical points of the theory, while nothing special was observed at 
$\mu_{\mathrm{crit}, 1\pm }$. In \cite{BMT:3D.mass.grav.Asymptotic.charges}, being aware of the fact that logarithmic 
modes typically appear when either $c_{-}$ or $c_{+}$ vanishes, the authors pointed out that it would be interesting to 
see whether the logarithmic solutions that EMG exhibits at $\mu_{\mathrm{crit}, 2 \pm }$ are also solutions at 
$\mu_{\mathrm{crit}, 1 \pm }$. We answer this question below. 

\begin{figure} [h!] 
\centering
{\includegraphics[width=12.5cm]{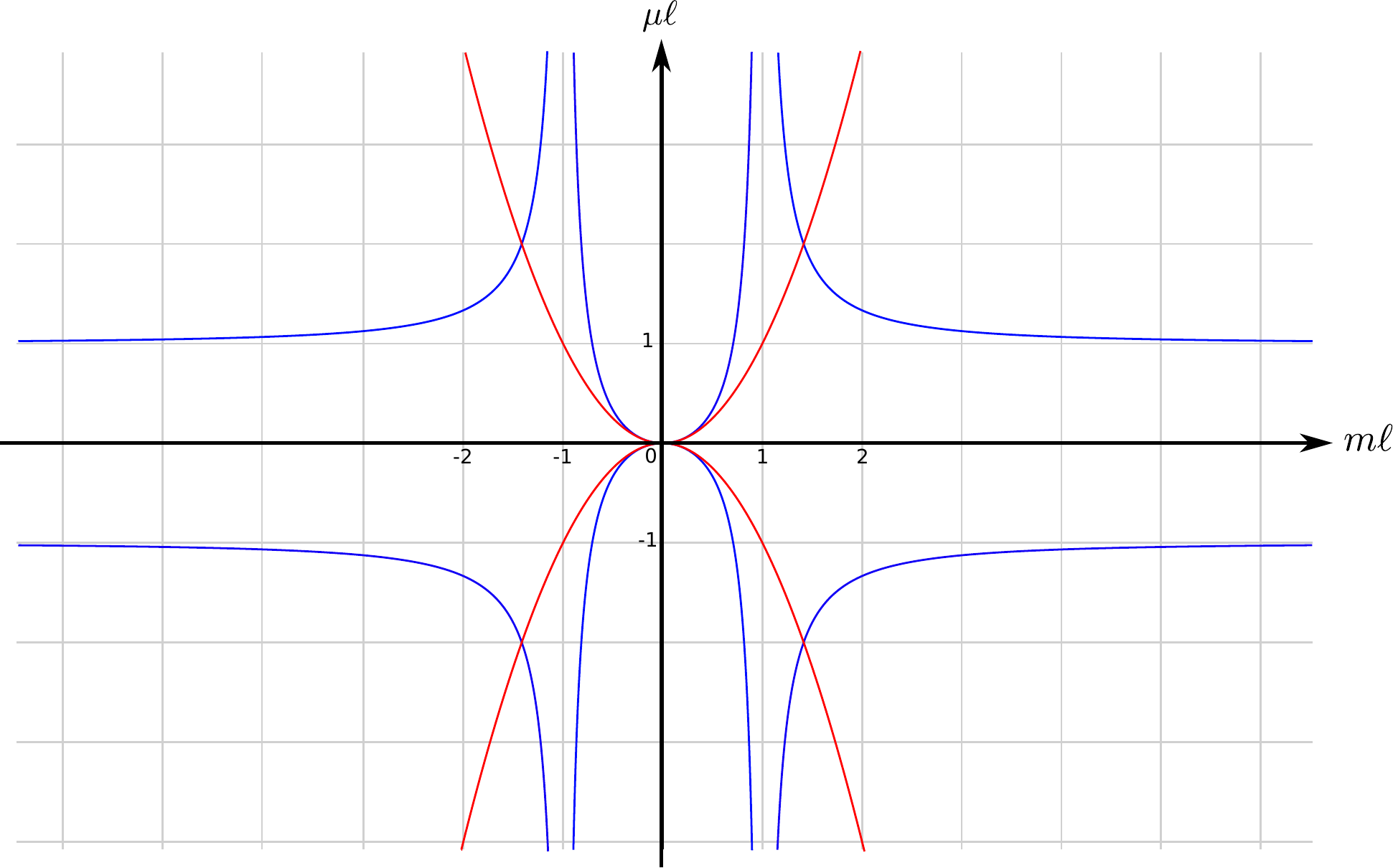}}
\caption{$\mu\ell$ (vertical axis) as a function of $m\ell$ (horizontal axis); $m^2>0$ is assumed here. Red curves correspond to $\mu = \mu_{\mathrm{crit}, 1 \pm }$, where $c_{\pm }=0$; blue curves correspond to $\mu = \mu_{\mathrm{crit}, 2 \pm }$, where $c_{\pm }=0$ too. In the large $| m\ell |$ limit one recovers the chiral points of TMG $\ell \mu_{\mathrm{crit}, 2 \pm } =\pm 1$. At $|m\ell |=1$ the critical value $\ell \mu_{\mathrm{crit}, 2 \pm }$ diverges and $c_-=c_+=0$. The case $m^2<0$ does not exhibit poles, and the behavior $\ell \mu_{\mathrm{crit}, 2 \pm } \to \pm1$ is recovered when $|m\ell |$ is large.} 
\end{figure}

\section{Exotic waves on black holes}
\label{sec:waves}

We will consider a generalization of the Ba\~nados-Teitelboim-Zanelli (BTZ) solution \cite{BTZ} that describes 
gravitational waves propagating on a stationary black hole geometry. This includes AdS-waves as a particular case. We 
will study systematically all the possible long distance behavior of such solutions near the AdS$_3$ boundary, showing 
that no log-modes appear at $\mu_{\mathrm{crit}, 1}$. 

\subsection{Deformed BTZ geometry}

Let us start by considering the extremal BTZ metric with mass parameter $M$ and spin parameter $J = \mp \, M\ell$; namely
\begin{equation}
\label{eq:ebtz}
	d s_{M, \mp \ell M}^{2} = -\dfrac{\left(r^{2}-r_{+}^{2}\right)^{2}}{\ell^{2} r^{2}} d t^{2} + \dfrac{\ell^{2} r^{2}}{\left(r^{2}-r_{+}^{2}\right)^{2}} d r^{2} + \dfrac{r^{2}}{\ell^{2}}\left(\ell d \varphi \pm \dfrac{r_{+}^{2}}{r^{2}} d t\right)^{2} \,,
\end{equation}
where $t\in \mathbb{R}$, $\varphi\in [0,2\pi ]$, $r\in \mathbb{R}_{\geq 0}$, and where $r_{+}=4GM\ell $ is the black hole horizon, with $G$ being the Newton constant. Now, perform a deformation of 
\eqref{eq:ebtz} of the type\footnote{Similarly, one may consider the extremal BTZ solution with $J=+\ell M$ deformed with a piece $h(x^-,r)(dx^{-})^2$.}
\begin{equation}
\label{eq:deformation}
	d s^{2} \equiv g_{\mu \nu} d x^{\mu} d x^{\nu} = d s_{M,-\ell M}^{2} + h(x^{+}, r)(dx^{+})^{2} \,,
\end{equation}
where $x^{\pm } = t \pm \ell\varphi$. In order to solve EMG field equations, a wave profile of the form $h(x^{+}, r) = 
f(x^{+}) \, F(r)$ must satisfy the following linear equation
\begin{equation}
\label{eq:Fdiffeq1}
	\left( p_1(r) \, \dfrac{d}{dr} + p_2(r) \, \dfrac{d^2}{dr^2} + p_3(r) \, \dfrac{d^3}{dr^3} + p_4(r) \, \dfrac{d^4}{dr^4} \right) \,F(r) = 0 \,,
\end{equation}
where the coefficients $p_i(r)$ are given by
\begin{equation}
\label{eq:Fdiffeq2}
\begin{split}
	p_1(r) &= -\frac12 \dfrac{r^2-r_+^2}{r^3} + \dfrac{3}{2}\dfrac{r_+^2(r^2-r_+^2)}{\mu\ell r^5} - \dfrac{3}{2} \dfrac{r_+^2(r^2-r_+^2)(2r^2-5r_+^2)}{m^2\ell^2r^7} \,, \\
	p_2(r) &= \frac12 \dfrac{r^2-r_+^2}{r^2} - \dfrac{3}{2}\dfrac{r_+^2(r^2-r_+^2)}{\mu\ell r^4} + \dfrac{3}{2} \dfrac{r_+^2(r^2-r_+^2)(2r^2-5r_+^2)}{m^2\ell^2r^6} \,, \\
	p_3(r) &= - \dfrac{1}{2}\dfrac{(r^2-r_+^2)^2}{\mu\ell r^3} - \dfrac{(r^2-r_+^2)^2(r^2+3r_+^2)}{m^2\ell^2r^5} \,, \\
	p_4(r) &= -\frac12 \dfrac{(r^2-r_+^2)^3}{m^2\ell^2 r^4} \,.
\end{split}
\end{equation}
For generic values of the coefficients, such an equation has solutions of the form $F(r) = A \: (r^2-r_+^2)^\Delta$, leading to the  
the indicial polynomial
\begin{equation}
\label{eq:indicial}
	\Delta(\Delta-1)\left[ \Delta^2 - \left( 1-\dfrac{m^2\ell}{2\mu} \right)\,\Delta + \frac14 - \dfrac{m^2\ell^2}{4} - \dfrac{m^2\ell}{4\mu} \right] = 0 \,,
\end{equation}
which, generically, has four different roots $\Delta = \left\lbrace 0 \,, 1 \,, \Delta_- \,, \Delta_+ \right\rbrace $, with
\begin{equation}
\label{eq:general.exponents}
	\Delta_{\pm} = \frac12 - \dfrac{m^2\ell}{4\mu} \pm \dfrac{m^2\ell}{4\mu} \sqrt{1+\dfrac{4\mu^2}{m^2}} \,.
\end{equation}
Then, the complete solution for the wave profile $h(x^+,r)$ reads
\begin{equation}
\label{eq:general.deformation}
	h(x^{+}, r) = D(x^{+}) + C(x^{+})(r^2-r_+^2) + B(x^{+})(r^2-r_+^2)^{\Delta_+} + A(x^{+})(r^2-r_+^2)^{\Delta_-} \,,
\end{equation}
with $A(x^{+})\,, B(x^{+})\,, C(x^{+})\,, D(x^{+})$ being four arbitrary functions that depend only on $x^{+}$; that is, 
$\partial_{-}A(x^{+})=0 \,, \partial_{-}B(x^{+})=0 \,, \partial_{-}C(x^{+})=0 \,, \partial_{-}D(x^{+})=0$. The constant and quadratic terms in (\ref{eq:general.deformation}), corresponding to $\Delta =0$ and $\Delta =1$ respectively, can be removed by local diffeomorphisms 
\cite{AyonBeato:NMG.AdSwaves}, i.e. they are solutions that are already present in 3D Einstein theory. In contrast, the modes $\Delta_{\pm}$ correspond to massive modes of EMG and are associated to the local degrees of freedom of the theory. Solution (\ref{eq:general.deformation}) generalizes the solutions found in \cite{Giribet.Oliva:EMG.More.Vacua} at the chiral points $\mu = \mu_{\rm crit,2\pm }$. Geometry (\ref{eq:deformation}) with (\ref{eq:general.exponents})-(\ref{eq:general.deformation}) represents a gravitational wave on  an extremal black hole. The wave co-rotates with the black hole, having an off-diagonal term 
\begin{equation}
g_{\varphi t}=\frac{r_+^2}{\ell} + \frac{\ell }{2} \, h(x^+ ,r).
\end{equation}
Assuming $\Delta_+ > 0$  and $\Delta_- < 0$, this includes an asymptotically AdS$_3$ solution like
\begin{equation}
ds^2 = \left( \frac{2r_+^2}{{\ell }} + \ell A(t+\ell\varphi ) \, (r^2-r_+^2)^{\Delta_-}\right)\, d\varphi \, dt + \, \ldots
\end{equation}
where the ellipsis stand for the diagonal terms. This moves clockwise as the black hole. The geometry is wound around the horizon and the effect of the deformation $h(x^+ ,r)$ get diluted near the boundary. The full geometry has scalar invariants 
\begin{equation}
\text{Tr} ( \text{Ric}^n)\equiv R_{\mu_1}^{\, \, \mu_2} R_{\mu_2}^{\, \, \mu_3} \, \ldots \, R_{\mu_{n}}^{\, \, \mu_1}     = -\frac{6(-2)^{n-1}}{\ell^{2n}}\, , \ \
\end{equation} 
which are those of AdS$_3$ space, although it is not locally equivalent to AdS$_3$. In fact, for $A(x^+)\neq 0$ or $B(x^+)\neq 0$, the geometry is not conformally flat.  

In conclusion, the propagating waves (\ref{eq:deformation}) can be seen as a fully backreacting, massive excitation of the black hole background. To reinforce this interpretation, we notice that a perturbation of the form $\phi(t,\varphi,r) = f(x^+ )  (r^2-r_+^2)^{\Delta }$ satisfies the wave equation
\begin{equation}
	\Box \left( \Box - K_{\Delta } \right) \phi(t, \varphi, r) = 0 \,, \label{Gesta}
\end{equation}
where $\Box $ is the d'Alembert operator of the full deformed geometry (\ref{eq:deformation}), and where the effective mass $K_{\Delta }$ is
\begin{equation}
	K_{\Delta } = \dfrac{4}{\ell^2} \Delta (\Delta +1) \,,
\end{equation}
which, taking into account \eqref{eq:general.exponents}, reads
\begin{equation}
	K_{\Delta } = \dfrac{3}{\ell^2} + \dfrac{m^2\ell (m^2\ell -4\mu+2\mu^2\ell) }{2\mu^2\ell^2} \pm 
	\dfrac{(4\mu-m^2\ell)m^2\ell}{2\mu^2\ell^2} \sqrt{1+ \dfrac{4\mu^2}{m^2}} \,.
\end{equation}

For generic $K_{\Delta} \neq 0$, the space of solutions to the wave equation (\ref{Gesta}) is the direct sum of the kernels 
$\text{Ker}(\Box -K_{\Delta })+\text{Ker}(\Box )$. At the chiral points $\mu = \mu_{\rm crit,2-}$, $K_{\Delta}$ vanishes, two roots of the 
indicial polynomial become zero, and a new logarithmic solution to (\ref{Gesta}) appears. This logarithmic solution is in the difference 
of kernels $\text{Ker}(\Box^2)-\text{Ker}(\Box )$, as usual with confluent differential equations.


Below, we will discuss systematically the different confluent points of the wave equation to see where logarithmic solutions 
actually occur. To organize the discussion, we will classify the confluent points in terms of their degree of degeneracy of the 
roots of \eqref{eq:indicial}. The different cases are: (a) the roots $\Delta_{\pm}$ collide, that is 
$\Delta_+ = \Delta_- $; (b) one of the roots $\Delta_{\pm}$ goes to either 0 or 1; (c) the limiting case where both 
$\Delta_{-}$ and $\Delta_{+}$ coincide with either 0 or 1. The case (a) occurs where $m^2=-4\mu^2$ and we will refer to it as the 
`degenerate point' or the `confluent point'. The case (b) corresponds to the chiral point $\mu = \mu_{\rm 
crit,2\pm}$ and, therefore, we will refer
to it as the `chiral point'. The case (c), to which we will refer as the `double confluent point', happens when $\mu = \mu_{\rm crit,2\pm} = \pm 1/(2\ell)$ . Finally, we will analyze the point $\mu = \mu_{\rm crit,1\pm}$, for which no special behavior is observed.

\subsection{Topologically massive gravity}

Let us start by studying the TMG limit of the general solutions, which corresponds to $m^2 \rightarrow \infty$. In this limit, the exponents reduce to $\Delta_+ = (1+\mu\ell)/2$ and $\Delta_{-} \rightarrow -\infty$, yielding
\begin{equation}
\label{eq:TMG.general.solution}
	h(x^{+}, r) = B(x^{+})(r^2-r_+^2)^{(1+\mu\ell)/2} \,,
\end{equation}
together with the modes $\Delta=0$ and $\Delta=1$ of GR. These are a generalization of the so-called AdS-waves of TMG \cite{AyonBeato:NMG.AdSwaves, GAY:0811}. For $B = {\rm const}$, solution (\ref{eq:TMG.general.solution}) is a stationary deformation of the BTZ black hole; and it is worth mentioning that this is not in contradiction with the Birkhoff-like theorems known for TMG
\cite{Aliev.Nutku:TMG.Birkhoff, Cavaglia:TMG.Birkhoff}. In particular, the existence of solution (\ref{eq:TMG.general.solution}) is consistent with a conjecture in \cite{Chiral.Gravity.Log}, which states that, at
the chiral point $\mu \ell =-1$, all stationary TMG solutions that satisfy the Brown-Henneaux boundary conditions are Einstein manifolds, cf. \cite{Compere:TMG.NonEinstein}. In fact, when $\mu \ell =-1$, the solution can be seen to be a solution of GR. For $\mu\ell < -1$, the deformation satisfies the Brown-Henneaux asymptotic
boundary conditions \cite{Brown.Henneaux:Central.Charge} and so it represents an asymptotically AdS$_3$ non-Einstein space. The case $\mu =0$ is special as the geometry with $h(x^{+}, r) = B(x^{+})(r^2-r_+^2)^{1/2}$ turns out to be conformally flat without being an Einstein manifold, so it corresponds to a non-trivial solution of 3D conformal gravity exhibiting the typical linear behavior $h \sim r $ at large distance.

\subsection{Confluent points}
\label{sec:degenerate}
When $m^2=-4\mu^2$, the roots $\Delta_{\pm}$ collide, i.e.
\begin{equation}
	\Delta_{+} =\Delta_{-} = \dfrac{1}{2}+\mu \ell \,.
\end{equation}
Since these solutions coincide, a new linearly independent solution to \eqref{eq:Fdiffeq1}-\eqref{eq:Fdiffeq2} 
must emerge. As probably expected, this new solution has a logarithmic behavior; more precisely
\begin{equation}
\label{eq:def.degenerate}
	h(x^{+},r) = B(x^{+})(r^2-r_+^2)^{{1}/{2}+\mu \ell } + A(x^{+})(r^2-r_+^2)^{{1}/{2}+\mu \ell }\log(r^2-r_+^2) \,.
\end{equation}
Depending on whether $\mu \ell <-1/2$ or $\mu \ell >-1/2$, the function $h(x^+,r)$ that controls the deformation of the BTZ 
geometry diverges at either the black hole horizon $r=r_+$ or at the boundary $r=\infty $, respectively.
At infinity, the behavior may actually be regular: The solution corresponding to the $B$-mode in (\ref{eq:def.degenerate}) obeys 
the Brown-Henneaux boundary condition if $\mu \ell < -1/2$.

\subsection{Chiral points}
\label{sec:critical}

More relevant for our discussion are the points $\mu = \mu_{\rm crit, 2 \pm} $, where one of the roots $\Delta_{\pm}$ of the indicial polynomial degenerates to either 0 or 1 and where one 
of the central charges \eqref{eq:central.charges} vanishes. There, again, new solutions to \eqref{eq:Fdiffeq1}-\eqref{eq:Fdiffeq2} that involve logarithms appear. Such solutions to EMG were already studied in ref.~\cite{Giribet.Oliva:EMG.More.Vacua}. They can be of two types: 

First, consider $\mu =\mu_{\rm crit, 2 +}$, which yields $c_+=0$. At this point, $\Delta_+ =1$ and the deformation \eqref{eq:deformation} takes the form
\begin{equation}
\label{eq:def.critical2p}
	h(x^{+},r) = B(x^{+})(r^2-r_+^2)^{(1-m^2\ell^2)/2} + A(x^{+})(r^2-r_+^2) \log(r^2-r_+^2) \,.
\end{equation}
For $m^2\ell^2\geq 1$, the $B$-mode of (\ref{eq:def.critical2p}) respects the Brown-Henneaux boundary conditions and so it gives an asymptotically AdS$_3$ solution. The $A$-mode, in contrast, neither respect the strong \cite{Brown.Henneaux:Central.Charge} nor the weakened \cite{Grumiller.Johansson:TMG.log.conditions} asymptotically AdS$_3$ boundary conditions. 

Second, consider the other chiral point, namely $\mu = \mu_{\rm crit, 2 -}$. In this case, $\Delta _-=0$ and, again, a new logarithmic mode appears. In this case, the wave 
profile $h(x^{+},r)$ takes the form\footnote{Logarithmic solutions like (\ref{eq:def.critical2p})-(\ref{eq:def.chiral2m}) were shown to exist in TMG, NMG, MMG, and in theories such as Zwei Dreibein Gravity (ZDG) \cite{ZDG} whenever one of the central charges of the dual CFT$_2$ vanishes \cite{Bergshoeff:1401}.}
\begin{equation}
\label{eq:def.chiral2m}
	h(x^{+},r) = B(x^{+})(r^2-r_+^2)^{(1+m^2\ell^2)/2} + A(x^{+}) \log(r^2-r_+^2) \,.
\end{equation}
At $\mu = \mu_{\rm crit, 2 -}$, one finds $c_-=0$ and $K_{\Delta }=0$. In fact, the long range mode, namely the logarithmic mode in (\ref{eq:def.chiral2m}), is interpreted as appearing due to the massless graviton: This logarithmic solution belongs to $\text{Ker}(\Box^2)-\text{Ker}(\Box )$. 

The observation that the effective mass $K_{\Delta }$ of the wave equation (\ref{Gesta}) vanishes should not be mistaken for statement that solution (\ref{eq:def.chiral2m}) has vanishing mass. In fact, it is not the case: The $A$-mode in (\ref{eq:def.chiral2m}) respects the boundary conditions of \cite{Grumiller.Johansson:TMG.log.conditions} and so it can be thought of as a solution\footnote{The $B$-mode, on the other hand, respects the stronger (Brown-Henneaux) AdS$_3$ boundary conditions provided $m^2\ell^2\leq -1$.} in AdS$_3$ with non-vanishing mass. Its mass, according to the computation in \cite{Giribet.Oliva:EMG.More.Vacua}, is given by
\begin{equation}
M=\frac{1}{4\pi G\ell }\Big( 1+\frac{1}{m^2\ell^2}\Big) \int_{0}^{2\pi \ell }A(\tau ) \, d\tau \,.
\end{equation}

\subsection{Double confluent points}
\label{sec:coincident}
Now, let us study the `double confluent points', which correspond to $\mu_{\rm crit, 2\pm} = \pm 1/(2\ell)$, where three 
roots of the indicial polynomial \eqref{eq:indicial} coincide: 

At the point $\mu=\mu_{\rm crit, 2+} = + 1/(2\ell)$, one finds $\Delta_{+} = \Delta_{-} =  1$ and $c_+=0$. The wave profile $h(x^{+},r)$ in this case takes the form
\begin{equation}
\label{eq:coincidentp}
	h(x^{+},r) = B(x^{+})(r^2-r_+^2) \log(r^2-r_+^2) + A(x^{+}) (r^2-r_+^2)\log^2(r^2-r_+^2)  \,.
\end{equation}
This type of $h \sim \log^2(r)$ solutions also appear in other higher-order generalization of TMG; see for instance \cite{AyonBeato:NMG.AdSwaves}.

At $\mu=\mu_{\rm crit, 2-} = - 1/(2\ell)$, on the other hand, one finds $\Delta_{+} = \Delta_{-} = 0$ and $c_-=0$, and, just as in the previous case, two new 
logarithmic modes appears; namely
\begin{equation}
\label{eq:coincidentm}
	h(x^{+},r) = B(x^{+})\log(r^2-r_+^2) + A(x^{+}) \log^2(r^2-r_+^2)  \,,
\end{equation}
which, for $A=0$, turns out to be asymptotically AdS$_3$ in the sense of \cite{Grumiller.Johansson:TMG.log.conditions}.

\subsection{Are any other critical points?}
\label{sec:critical1}

Now, let us analyze the points $\mu_{\rm crit, 1 \pm} = \pm m^2\ell$, where the charges (\ref{eq:central.charges}) obtained in \cite{BMT:3D.mass.grav.Asymptotic.charges} also vanish. At 
those points, no logarithmic behavior near the boundary of AdS$_3$ seems to occur. Consistently, $K_{\Delta }$ does not vanish 
there. This answers the question raised in \cite{BMT:3D.mass.grav.Asymptotic.charges} about 
the existence of such log-solutions at $\mu=\mu_{\rm crit, 1 \pm}$. 

Actually, it is easy to see from \eqref{eq:general.exponents} that at $\mu = \mu_{\rm crit, 1 +}$ one has 
$\Delta_{\pm} = (1 \pm \sqrt{1+4m^2\ell^2})/4$ whereas at $\mu = \mu_{\rm crit, 1 -}$ one has $\Delta_{\pm} = (3 \mp \sqrt{1+4m^2\ell^2})/4$, and so the solution takes the power-like form
\begin{equation}
	h(x^{+}, r) = B(x^{+})(r^2-r_+^2)^{\Delta_{+}} + A(x^{+})(r^2-r_+^2)^{\Delta_{-}} \,,
\end{equation}
with no logarithmic behavior. This might seem puzzling because, as we said, in any other massive deformation of 3D gravity that had been explored, whenever a 
central charge of the dual CFT$_2$ vanishes log-modes were shown to appear; this happens, for example, in TMG, NMG, MMG, ZDG. This invites us to return to the question about the discrepancy between (\ref{eq:central.charges}) and the central 
charges obtained in \cite{Giribet.Oliva:EMG.More.Vacua}, the later being non-zero at $\mu=\mu_{\rm crit, 1 \pm}$.

\section{Discussion}

A first observation to understand the reason for the discrepancy of the central charges computed in 
\cite{BMT:3D.mass.grav.Asymptotic.charges} and those computed in \cite{Giribet.Oliva:EMG.More.Vacua} is that, when taking the limit 
$\mu\to \infty$ in (\ref{eq:chiral.curves}) and, after that, taking the limit $m\to\infty$, one obtains $c_+=-c_- = 
3\ell/(2G)$, which agrees with the central charges of the so-called Exotic Gravity (EG) \cite{EG}. In contrast, if considers the result of \cite{Giribet.Oliva:EMG.More.Vacua} and takes the limit $\mu\to \infty$, $m\to\infty$ of that then one obtains 
$c_+ = c_- = 3\ell/(2G)$, which is the Brown-Henneaux central charge of Einstein gravity 
\cite{Brown.Henneaux:Central.Charge}. This means that the discrepancy between the charges can be traced back to the difference of the theories that are being considered: The papers \cite{BMT:3D.mass.grav.Asymptotic.charges} and \cite{Giribet.Oliva:EMG.More.Vacua} are actually dealing with different theories; while \cite{BMT:3D.mass.grav.Asymptotic.charges} deals with a 
Chern-Simons like computation in the higher-order extension of the EG \cite{EG}, \cite{Giribet.Oliva:EMG.More.Vacua} deals 
with an Abbott-Deser-Tekin (ADT) like computation\footnote{In \cite{BMT:3D.mass.grav.Asymptotic.charges}, it is affirmed that 
the main reason for the disagreement with \cite{Giribet.Oliva:EMG.More.Vacua} is that in the latter paper the ADT method 
is applied in the metric formulation while the stress-tensor is not the linearized limit of any consistent source tensor 
for the full EMG equations. However, since the linearized field equations of EMG are on-shell divergenceless, their 
contraction with a Killing vector of the background leads to a well-defined ADT-like conserved current.} of the 
higher-order extension of GR. Therefore, it should not come to a surprise that the charges do not agree. To understand 
this better, let us recall how it works in the case of the undeformed theory $\mu=\infty$, $m=\infty$: While the GR 
Lagrangian in terms of the vielbein 1-form $e^a=e^a_{\mu}dx^{\mu}$ and the spin connection 1-form 
$\omega^{ab}=\omega_{c}\epsilon^{abc}=\omega^{ab}_{\mu}dx^{\mu}$ reads $L_{\text{GR}}=\epsilon_{abc}(R^{ab}\wedge 
e^{c}+e^{a}\wedge e^{b}\wedge e^{c})$ with $R^{ab}$ being the curvature 2-form ($\ell=1$), the EG Lagrangian reads 
$L_{\text{EG}}=\epsilon_{abc}(\omega^{ab}\wedge d\omega^{c}+\frac 13
\omega^{a}\wedge\omega^{b}\wedge\omega^{c})+e_a T^a$ with $T^{a}$ being the torsion 2-form (see also \cite{ExoticCGM}). Both theories yield the same field equations (i.e. the 3D cosmological Einstein equations) but they yield different charges. Our interpretation is that the same is happening here with the massive deformations. 

Therefore, we think of the theory defined by the field equations (\ref{eq:eom})-(\ref{eq:HL.tensors}) at the chiral point $\mu=\mu_{\rm crit, 2 -}$ as the gravity dual of a CFT$_2$ with the central charges \cite{Giribet.Oliva:EMG.More.Vacua}
\begin{equation}
c_-=0 \,, \quad \quad c_+ = \frac{3\ell }{G}\Big(1-\frac{1}{m^2\ell^2}\Big) \,.
\end{equation}
At this point, the solutions of the theory exhibits the typical behavior $h\sim \log(r)$ of the Log-gravity \cite{Chiral.Gravity.Log}. This suggests that, at $\mu=\mu_{\rm crit, 2 -}$, provided one considers sufficiently weak AdS$_3$ boundary conditions, the dual CFT$_2$ turns out to be a logarithmic CFT$_2$ \cite{LCFT}; that is, a non-unitary CFT$_2$ whose Virasoro operators $L_0$ and $\bar{L}_0$ are not diagonalizable but form a Jordan block. This means that in the CFT$_2$ there exists a mixing between primary operators and other type of operators, called the logarithmic partners. In particular, the stress tensor may have a logarithmic partner with which it has a non-vanishing 2-point function. This mixing of the stress tensor and its partner is controlled by a new anomaly, $b$, which appears in the pole $\sim 1/z^4$ of the operator product expansion. In \cite{Grumiller:Shortcut.newanomalies}, a simple method to compute this anomaly for the case of a logarithmic CFT$_2$ with a massive AdS$_3$ gravity dual was given. Applying this method in the case of EMG, we find
\begin{equation}
b= -\dfrac{3\ell}{G}\Big(1+\frac{1}{m^2\ell^2}\Big) \,. \label{HHH}
\end{equation}
As a consistency check, we observe that in the limit $m\to \infty$, $\mu_{\rm crit, 2 -}$ tends to $-1/\ell$, which is the chiral point of TMG; in that limit, $b$ tends to $-c_+=-{3\ell}/{G}$, which is in perfect agreement with the anomaly coefficient of the Log-Gravity of TMG \cite{Chiral.Gravity.Log, Grumiller:Shortcut.newanomalies}.

\acknowledgments

The authors thank Mariano Chernicoff, Nicol\'as Grandi, Robert Mann and S.~N.~Sajadi for discussions. The work of G.G. has been partially supported by CONICET through a grant PIP 1109 (2017). J.O. is partially funded by the grant FONDECYT 1181047.

\end{document}